\pdfoutput=1
\newcommand*{\EXTENDEDVERSION}{}

\documentclass[10pt,conference]{IEEEtran}

\usepackage[utf8]{inputenc}
\usepackage{booktabs}
\usepackage{subcaption}
\usepackage{amsmath}
\usepackage{amssymb}
\usepackage{textcomp}
\usepackage[svgnames,table]{xcolor}
\usepackage{pgfpages}
\usepackage{pgfplots}
\usepackage{pgfplotstable}
\usepackage{tikz}
\usepackage{enumerate}
\usepackage{algorithmicx}
\usepackage{algorithm}
\usepackage[noend]{algpseudocode}
\usepackage{listings}
\usepackage{tabularx}
\usepackage{url}
\usepackage[multiple]{footmisc}
\usepackage[font=bf]{caption}
\usepackage{multirow}
\usepackage{arydshln}
\usepackage{xspace}
\usepackage{flushend}
\usepackage{diagbox}
\usepackage{bm}
\usepackage{siunitx}
\usepackage{multirow}
\ifdefined\EXTENDEDVERSION
  \usepackage[breaklinks,hidelinks,pdftitle={Differentially
  Testing Soundness and Precision of Program Analyzers},pdfauthor={Christian Klinger,
  Maria Christakis, Valentin Wuestholz},pdfkeywords={differential testing, program
  analysis, soundness, precision}]{hyperref}
\else \usepackage[breaklinks,hidelinks]{hyperref} \fi
\definecolor{light-gray}{gray}{0.75}
\definecolor{BrickRed}{rgb}{0.8, 0.25, 0.33}
\definecolor{lightorange}{HTML}{FFB74D}

\makeatletter
\newenvironment{btHighlight}[1][]
{\begingroup\tikzset{bt@Highlight@par/.style={#1}}\begin{lrbox}{\@tempboxa}}
{\end{lrbox}\bt@HL@box[bt@Highlight@par]{\@tempboxa}\endgroup}

\newcommand\btHL[1][]{%
  \begin{btHighlight}[#1]\bgroup\aftergroup\bt@HL@endenv%
}
\def\bt@HL@endenv{%
  \end{btHighlight}%
  \egroup
}
\newcommand{\bt@HL@box}[2][]{%
  \tikz[#1]{%
    \pgfpathrectangle{\pgfpoint{0.3pt}{0pt}}{\pgfpoint{\wd #2}{\ht #2}}%
    \pgfusepath{use as bounding box}%
    \node[anchor=base west,fill=lightorange,outer sep=0pt,inner xsep=0.3pt,inner ysep=0pt,minimum height=\ht\strutbox+0.3pt,#1]{\raisebox{0.3pt}{\strut}\strut\usebox{#2}};
  }%
}
\makeatother
\usetikzlibrary{calc,trees,positioning,arrows,chains,shapes.geometric,%
  decorations.pathreplacing,decorations.pathmorphing,decorations.text,shapes,%
  matrix,shapes.symbols,patterns,shadows}
\pgfplotsset{compat=newest}
\usepgfplotslibrary{
  colorbrewer,
  groupplots
}
\newcommand*\Let[2]{\State #1 $\gets$ #2}

\newcommand*\Fcall[1]{\textsc{#1}}

\algrenewcommand\alglinenumber[1]{\tiny\color{Black!70}{#1}}
\algrenewcommand\algorithmicforall[2]{\textbf{for} $i=$ #1 \textbf{to} #2}
\algrenewtext{ForAll}{\algorithmicforall}
\algnewcommand\algorithmicswitch{\textbf{switch}}
\algnewcommand\algorithmiccase{\textbf{case}}
\algdef{SE}[SWITCH]{Switch}{EndSwitch}[1]{\algorithmicswitch\ #1\ \algorithmicdo}{\algorithmicend\ \algorithmicswitch}%
\algdef{SE}[CASE]{Case}{EndCase}[1]{\algorithmiccase\ #1}{\algorithmicend\ \algorithmiccase}%
\algtext*{EndSwitch}%
\algtext*{EndCase}%
\algdef{SE}[SUBALG]{Indent}{EndIndent}{}{\algorithmicend\ }%
\algtext*{Indent}
\algtext*{EndIndent}
\lstdefinestyle{basic}{%
  morekeywords     = [1]{var},%
  morekeywords     = [2]{assert, assume},%
  keywordstyle     = \bfseries\color{DarkBlue},%
  keywordstyle     = [2]\bfseries\color{BrickRed},
  commentstyle     = \ttfamily\color{Black!70}\lst@ifdisplaystyle\footnotesize\fi,%
  basicstyle       = \ttfamily\lst@ifdisplaystyle\footnotesize\fi,%
  columns          = [c]fixed,%
  aboveskip        = 0mm,%
  belowskip        = 2mm,%
  keepspaces       = true,%
  mathescape       = true,%
  escapechar       = ?,%
  tabsize          = 2,%
  numbers          = left,%
  numberstyle      = \tiny\color{Black!80},%
  numbersep        = 1.0em,%
  stepnumber       = 1,%
  firstnumber      = 1,%
  showstringspaces = false,%
  captionpos       = b,%
  extendedchars    = true,%
  xleftmargin      = 2.5em,%
  upquote          = true,%
  abovecaptionskip = 0.5em,%
  belowcaptionskip = 0.5em,%
  moredelim        = **[is][{\btHL[fill=light-gray]}]{°}{°},%
}

\lstdefinestyle{clang}{%
  language         = C,%
  style            = basic,%
}
\pgfplotsset{width=7cm,compat=newest}
\usepgfplotslibrary{colorbrewer}

\newcommand\code[1]{\lstinline[style=clang]{#1}}

\renewcommand{\figurename}{Figure}
\renewcommand{\tablename}{Table}
\renewcommand{\thetable}{\arabic{table}}

\newcommand\secref[1]{Sect.~\ref{#1}}

\newcommand\figref[1]{Fig.~\ref{#1}}

\newcommand\algoref[1]{Alg.~\ref{#1}}
\newcommand\tabref[1]{Tab.~\ref{#1}}

\newcommand\tool{$\alpha$-Diff\xspace}

\makeatletter
\newif\if@anonymize

\@anonymizefalse % Uncomment for camera-ready version

\if@anonymize
  \newcommand\anonymize[1]{Link removed for double-blind review.}
\else
  \newcommand\anonymize[1]{\tiny#1}
\fi
\makeatother

\newcommand{\maxhl}[1]{{\cellcolor[gray]{0.8}} #1}

\newcommand{\nd}{$\star$}
\begin{document}

\bstctlcite{IEEEexample:BSTcontrol}

%% Title
\title{Differentially Testing\\Soundness and Precision of Program Analyzers}

\author{
\IEEEauthorblockN{Christian Klinger}
\IEEEauthorblockA{\textit{Saarland University, Germany}\\
io@klinch.de}
\and
\IEEEauthorblockN{Maria Christakis}
\IEEEauthorblockA{\textit{MPI-SWS, Germany}\\
maria@mpi-sws.org}
\and
\IEEEauthorblockN{Valentin W{\"u}stholz}
\IEEEauthorblockA{\textit{ConsenSys, Germany}\\
valentin.wustholz@consensys.net}
}

\maketitle

\begin{abstract}
  In the last decades,
  numerous program analyzers have been developed both by academia and
  industry. Despite their abundance however, there is currently no
  systematic way of comparing the effectiveness of different analyzers
  on arbitrary code.  In this paper, we present the first automated
  technique for differentially testing soundness and precision of
  program analyzers. We used our technique to compare six mature,
  state-of-the art analyzers on tens of thousands of automatically
  generated benchmarks. Our technique detected soundness and precision
  issues in most analyzers, and we evaluated the implications of these
  issues to both designers and users of program analyzers.
\end{abstract}

\vspace{0.5em}

%% ----------------------------------------------------------------
\section{Introduction}
\label{sec:Introduction}
%% ----------------------------------------------------------------

Program analyzers are very effective in detecting code issues and,
especially in recent years, they are increasingly applied in industry
to detect defects in real-world software. Notable examples of such
analyzers are Facebook's
Infer~\cite{CalcagnoDistefano2015,CalcagnoDistefano2011}, which is
used to find resource leaks and null-pointer exceptions in Android and
Java applications, Microsoft's SAGE~\cite{GodefroidLevin2008}, which
is credited to have detected about one third of all security
bugs that were discovered with fuzzers during the development
of Windows~7~\cite{BounimovaGodefroid2013}, and AbsInt's
Astr\'ee~\cite{BlanchetCousot2003,CousotCousot2005}, which can prove
the absence of runtime errors and invalid concurrent behavior in
safety-critical software, such as that of Airbus.

Despite the abundance of program analyzers that have been produced
both by academia and industry, there is currently no systematic way of
comparing their effectiveness on arbitrary code. To compare the
soundness and precision of a set of analyzers, one could try them on a
number of programs to get a feel for their false positive or false
negative rates. However, just classifying the generated warnings as
false or true positives would require considerable human effort, let
alone determining whether any bugs are missed.

Alternatively, one could rely on the outcome of software verification
competitions such as SV-COMP~\cite{SV-COMP}, which compares program
analyzers based on an established collection of verification tasks.
Although verification competitions are extremely valuable in
increasing the visibility of new analyzers, providing a discussion
forum for state-of-the-art techniques, and maintaining a set of
programs with explicit properties to be checked, the verification
tasks are rather stable. As a consequence, program analyzers can be
designed to perform well in such competitions by specifically
tailoring their techniques to the given benchmarks. In addition, the
explicit checks might not be representative of properties in
real-world software.

\textbf{Our approach.} To address these issues, we present the first
automated technique for differentially testing program analyzers on
arbitrary code. Given a set of seed programs, our approach
automatically generates program-analysis benchmarks and compares the
soundness and precision of the analyzers on these benchmarks. As a
result, the effectiveness of the different program analyzers is
evaluated in a systematic and automated way, the benchmarks are hardly
predictable, and the explicit checks can be parameterized to express
several types of properties, for instance, numerical, non-nullness, or
points-to properties.

However, as for existing differential testing techniques, it is
challenging to automatically derive the ground truth, for example,
which warnings are indeed true positives or which errors are missed.
We address this challenge by leveraging Engler et al.’s
``bugs-as-deviant-behavior''
strategy~\cite{EnglerChen2001}. Specifically, when most program
analyzers agree that a certain property does not hold, our approach
detects a potential soundness issue in the deviant analyzers, which
find that the property does hold.  Conversely, we detect a potential
precision issue when a few analyzers claim that a property does not
hold, while the majority of analyzers verify the property.

The work most closely related to ours is by Kapus and Cadar, who use
random program generation and differential testing to find bugs in
symbolic execution engines~\cite{KapusCadar2017}. In contrast to this
work, our approach focuses on detecting soundness and precision issues
in any program analyzer, potentially including a test generator based
on symbolic execution. Moreover, our technique automatically generates
program-analysis benchmarks from a given set of seed programs,
possibly containing code that is difficult to handle by program
analysis. In general, it is very challenging to randomly generate
programs from scratch such that they reveal soundness and precision
issues in mature analyzers, which is why our technique leverages the
seed programs.

Overall, we expect our approach to guide users in making informed
choices when selecting a program analyzer. However, this is not to say
that the best analyzer is the most sound; users have varying needs
depending on how critical the correctness of their code is and where
in the software development cycle they are (e.g., before a code review
or product release)~\cite{ChristakisBird2016}. We also expect our
technique to assist designers of analyzers in detecting soundness and
precision issues of their implementation, and to help enrich the
collection of tasks used in verification competitions by automatically
generating challenging, yet less predictable, benchmarks.

\textbf{Contributions.} We make the following contributions:
\begin{itemize}
\item We present the first automated technique for differentially
  testing soundness and precision of program analyzers on arbitrary
  code.
\item We implement this technique in a tool architecture that compares
  analyzers on C programs and may be instantiated with any program
  analyzer for C.
\item We used our technique to compare six state-of-the-art program
  analyzers on about 26,000 programs. We found soundness and precision
  issues in four (out of six) analyzers, and we evaluated their
  significance to both program-analysis designers and users.
\end{itemize}

\textbf{Terminology.} Since terminology varies across different
program-analysis techniques, we introduce the following terms that we
use throughout the paper. \emph{Sound} program analysis
over-approximates the set of possible executions in a given program in
order to prove the absence of errors in the program. Due to its
over-approximation, sound analysis may generate \emph{false
  positives}, that is, spurious warnings about executions that are not
erroneous or even possible in the program. In contrast, a \emph{true
  positive} is a warning about an actual error in the program.
\emph{Unsound} program analysis may over-approximate certain program
executions and under-approximate others in order to find bugs, rather
than prove their absence. In case an unsound analysis fails to detect
an error in the program, we say that it generates a \emph{false
  negative}.

An \emph{imprecise} program analysis abstracts certain program
executions such that it considers more executions than those feasible
in the program. Although an imprecise analysis might generate false
positives, it is not necessarily sound. On the other hand, a
\emph{precise} program analysis uses abstractions that closely
describe executions in the program to generate as few false positives
as possible.

\textbf{Outline.} The next section gives an overview of our approach
through a running example.  \secref{sec:Technique} explains the
technical details of our approach, while \secref{sec:Implementation}
describes our implementation. We present our experimental evaluation
and threats to the validity of our results in
\secref{sec:Experiments}. We discuss related work in
\secref{sec:RelatedWork} and conclude in \secref{sec:Conclusion}.

%% ----------------------------------------------------------------
\section{Overview}
\label{sec:Overview}
%% ----------------------------------------------------------------

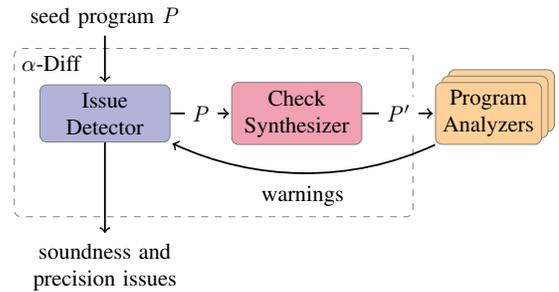
\begin{figure}[t]
\centering
\scalebox{0.85}{
  \begin{tikzpicture}[align=center, node distance=1.5cm]

    %% Nodes
    \node[draw=none] (P) at (0,0) {seed program $\mathit{P}$};
    \node[draw, rounded corners=3, fill=white, draw=black!50, below of=P, xshift=1.7cm, yshift=-0.3cm, dashed] (AD) {
    \begin{minipage}[t][2.4cm]{6.0cm}
    \tool
    \end{minipage}};
    \node[draw, rounded corners=3, fill=NavyBlue!30, draw=black!50, below of=P, yshift=0cm, text width=1.8cm] (F) {Issue Detector};
    \node[draw, rounded corners=3, fill=Crimson!40, draw=black!50, right of=F, xshift=1.5cm, text width=1.8cm] (X) {Check Synthesizer};
    \node[draw, rounded corners=3, fill=DarkOrange!40, draw=black!50, right of=X, xshift=1.5cm, double copy shadow={shadow xshift=0.1cm, shadow yshift=0.1cm}] (L) {Program\\Analyzers};
    \node[draw=none, below of=F, yshift=-0.9cm, text width=2.5cm] (T) {soundness and precision issues};
    %% Arrows
    \draw[thick, ->, shorten >=1pt] (P) -- (F);
    \draw[thick, ->, shorten >=1pt] (F) -- (T);
    \draw[thick, ->, shorten >=1pt] (F) -- (X) node[midway, anchor=center, fill=white] {$\mathit{P}$};
    \draw[thick, ->, shorten >=1pt] (X) -- (L) node[midway, anchor=center, fill=white] {$\mathit{P'}$};
    \draw[thick, ->, shorten >=1pt, out=-155, in=-25, looseness=0.9] (L.south west) to node[midway, anchor=center, fill=white, yshift=-0.35cm] {warnings} (F.south east);
  \end{tikzpicture}
}
\caption{Overview of the workflow and tool architecture.}
\label{fig:workflow}
\vspace{-1.5em}
\end{figure}

In this section, we illustrate the workflow and tool architecture of
our differential testing technique for program analyzers, shown in
\figref{fig:workflow}.

\textbf{Workflow.} Our technique, which is implemented in a tool
called \tool (pronounced ``alpha-diff''), takes as input one
or more correct seed programs that do not contain any (explicit or
implicit) assertions. Because these programs are both correct and
assertion-free, program analyzers (even if sound and imprecise) do not
generate any warnings for them.

Next, \tool parses a seed program, and based on one of its search
strategies (\secref{subsec:Synthesis}), a program location is
selected. At this location, \tool synthesizes and introduces a check,
in the form of an assertion, expressing a property of interest (e.g.,
a numerical property) and involving variables that are in scope at the
program location (\secref{subsec:Synthesis}). We call the resulting
program a variant of the seed program.

On this program variant, which now contains a single assertion, \tool
runs a set of program analyzers. The results of the analyzers, that
is, the presence or absence of any generated warnings for the
assertion, are recorded.

Subsequently, \tool selects a new program location in the same seed
program and repeats the process until a given \emph{budget} (i.e.,
number of synthesized checks for a particular seed program) is
depleted. The tool then continues to parse another seed program, if
any.

When all seed programs have been instrumented and analyzed until the
budget, \tool compares the recorded results and reports any
detected soundness and precision issues in the analyzers
(\secref{subsec:IssueDetection}).

\textbf{Example.} \figref{fig:smack} shows a simplified version of an
SV-COMP benchmark that we used in our experimental evaluation as a
seed program to test six program
analyzers. (Lines~\ref{line:soundness} and \ref{line:precision} should
be ignored for now.) Note that this program is correct and does not
contain any assertions. Therefore, when running all analyzers on this
code, no warnings are generated.

\begin{figure}[b]
\vspace{-1.5em}
\begin{lstlisting}[style=clang]
int main() {
  int i = 0;
  while (i < 100000) {
    °assert i != 13;° // soundness issue ?\label{line:soundness}?
    i = i + 1; ?\label{line:assignment}?
  }
  °assert i != 10;°   // precision issue ?\label{line:precision}?
  return i;
}
\end{lstlisting}
\vspace{-0.5em}
\caption{Soundness and precision issues in SMACK.}
\label{fig:smack}
\end{figure}

When passing this seed program to \tool, the `Check Synthesizer' (from
\figref{fig:workflow}), which is implemented to synthesize checks for
numerical properties, introduces the assertion on
line~\ref{line:soundness}. Our tool then runs all analyzers on the
resulting program variant, whose assertion can obviously fail. All
analyzers detect the assertion violation except for
CBMC~\cite{ClarkeKroening2004}, a bounded model checker for C and C++
programs, and SMACK~\cite{RakamaricEmmi2014}, a software verifier that
translates the LLVM intermediate representation into the Boogie
intermediate verification language~\cite{BarnettChang2005}. CBMC
typically unrolls loops as many times as necessary such that all bugs
are found, but we imposed a time limit on all analyzers, which proved
to be insufficient for CBMC to unroll the loop enough times such that
the assertion violation is detected. (Note, however, that CBMC does
find the bug with a higher time limit.) Still, CBMC soundly returns
`unknown', that is, no bugs were found but, due to reaching the time
limit, the program has not been verified. On the other hand, SMACK
claims to have verified the assertion on line~\ref{line:soundness},
which indicates a soundness issue.

We reported this
issue\footnote{\anonymize{\url{https://github.com/smackers/smack/issues/324}}}
to the designers of SMACK, who told us that the assertion violation is
missed due to a size-reduction heuristic, which searches for large
constants in SV-COMP benchmarks, such as \code{100000} in
\figref{fig:smack}, and replaces them with smaller numbers, in our
case with \code{10}, to reduce the benchmark size. This heuristic is
unsound and specifically tailored to the competition benchmarks.

In addition to being a source of unsoundness, this heuristic can also
be the cause of imprecision. For example, consider the
program variant of \figref{fig:smack} with the assertion on
line~\ref{line:precision} (and without the assertion on
line~\ref{line:soundness}). This assertion is introduced by \tool
within the budget that is assigned to the seed program from SV-COMP,
and it can never fail. All analyzers but SMACK are able to verify this
program. However, because of the size-reduction heuristic, SMACK knows
that variable \code{i} is equal to \code{10} right after the loop, and
therefore, the verifier reports an assertion violation, indicating a
precision issue.

%% ----------------------------------------------------------------
\section{Differential Testing of Program Analyzers}
\label{sec:Technique}
%% ----------------------------------------------------------------

In this section, we describe the main components of our workflow in
more detail and explain precisely how we detect soundness and
precision issues in program analyzers.

%% ----------------------------------------------------------------
\subsection{Check Synthesis}
\label{subsec:Synthesis}
%% ----------------------------------------------------------------

The check-synthesis component consists of two aspects: (1)~the
instrumentation, which creates a check and introduces it at a certain
location in the seed program to generate a variant, and (2)~the search
strategies, which explore the space of possible seed-program variants
that may be generated. We discuss these aspects next.

\textbf{Instrumentation.}
\algoref{alg:synthesis} describes how \tool generates a variant
$\mathit{P'}$ from a seed program $\mathit{P}$.
First, the algorithm selects a candidate expression $\mathit{e}$ at
program location $\mathit{l}$ in $\mathit{P}$. In our context, a
candidate expression is a pure expression of integral type that reads
from at least one variable, e.g., \code{c + 3} or \code{a[i]}, where
\code{c} is a variable of type \code{char} and \code{a} is an array of
integers. (Note that the choice of expression $\mathit{e}$ is made
according to a search strategy.)
Next, the algorithm generates a constant $\mathit{k}$, which is used,
together with expression $\mathit{e}$, to create an assertion of the
form \code{assert} $\mathit{e}$ \code{!=} $\mathit{k}$.
To generate variant $\mathit{P'}$, this assertion is inserted at
location $\mathit{l}$ in $\mathit{P}$. As an example, consider the
assertion on line~\ref{line:soundness} of \figref{fig:smack}, which is
introduced at the program location where expression \code{i}
(right-hand side of the assignment on line~\ref{line:assignment}) is
found in seed program $\mathit{P}$. Similarly, the assertion on
line~\ref{line:precision} is inserted at the location of the
\code{return} statement in $\mathit{P}$.

The assertions introduced by \tool in a seed program check numerical
properties. In particular, they check whether candidate expression
$\mathit{e}$ can ever have value $\mathit{k}$ at program location
$\mathit{l}$. If so, then the assertion can fail, and the tested
analyzers should detect this violation in order to be sound. If the
expression can never have this value at that location, then the
assertion cannot fail, and the analyzers should not detect any
violation in order to be precise.
%% For instance, the assertion on
%% line~\ref{line:soundness} of \figref{fig:smack} can fail since
%% \code{i} can have value \code{13} at that point in the code, whereas
%% on line~\ref{line:precision}, \code{i} can never be equal to
%% \code{10}, that is, the assertion cannot fail.

\begin{algorithm}[t]
  \caption{Check synthesis}
  \label{alg:synthesis}
  \hspace{0em}\textbf{Input:} seed program $\mathit{P}$
  \begin{algorithmic}[1]
    \small
      \State \textit{// Pick candidate expression $\mathit{e}$ at program location $\mathit{l}$.}
      \Let{$\mathit{e, l}$}{\Fcall{GetCandidateExpression}$(\mathit{P})$}
      \State \textit{// Pick constant $\mathit{k}$.}
      \Let{$\mathit{k}$}{\Fcall{GenerateConstant}$(\mathit{P}, \mathit{e})$}
      \State \textit{// Create check.}
      \Let{$\mathit{s}$}{\code{assert} $\mathit{e}$ \code{!=} $\mathit{k}$}
      \State \textit{// Instrument program $\mathit{P}$ by inserting statement $\mathit{s}$.}
      \Let{$\mathit{P'}$}{\Fcall{InsertAtLocation}$(\mathit{P}, l, s)$}
  \end{algorithmic}
  \hspace{0em}\textbf{Output:} program variant $\mathit{P'}$
\end{algorithm}

Note that, although \tool generates program variants that check
numerical properties, the `Check Synthesizer' of \figref{fig:workflow}
is configurable and may be extended to also synthesize other types of
properties. Still, as we discuss in \secref{subsec:Results}, numerical
properties were sufficient in detecting soundness and precision issues
in most of the analyzers we tested.

\textit{Batch checks.} To reduce the number of times the program
analyzers are invoked, \tool can also synthesize assertions with
multiple conjuncts, which we call `batch checks'. For example, a
program variant of \figref{fig:smack} could check whether \code{i} is
ever equal to \code{10} or \code{11} on line~\ref{line:precision} with
the assertion \code{assert i != 10 && i != 11}. Recall that, at this
program location, SMACK knows that \code{i} has value \code{10} and
would, therefore, detect a violation only due to the first conjunct of
the above assertion. For such cases, \tool uses divide-and-conquer to
eliminate conjuncts that do not cause any disagreement between the
program analyzers. We evaluate the effectiveness of our technique when
synthesizing assertions with batch checks in \secref{subsec:Results}.

\textbf{Search strategies.} In addition to generating a value
$\mathit{k}$, \algoref{alg:synthesis} also explores the search space
of possible candidate expressions $\mathit{e}$. Our technique
navigates this space using a number of \emph{static} and
\emph{dynamic} search strategies, which we describe below and evaluate
in \secref{subsec:Results}.
Note that the search strategies are applied by the
\textsc{GetCandidateExpression} function in \algoref{alg:synthesis},
which is configurable in our implementation.

\textit{Static strategies.} Static strategies traverse the abstract
syntax tree (AST) of the seed program to collect all the candidate
expressions. These strategies then compute a weight $w_e$ for each
candidate expression $e$ (according to a weight function), sum all
weights to compute the total $w_t$, and assign to each expression $e$
the probability $w_e / w_t$ of being selected by function
\textsc{GetCandidateExpression}. Overall, \tool supports three static
strategies that differ in their weight functions:
\begin{itemize}
  \item The \emph{Uniform-Random} strategy selects candidate
    expressions uniformly. In other words, all possible locations have
    a weight of $1$.
  \item The \emph{Breadth-Biased} strategy assigns to each candidate
    expression at location $\mathit{l}$ a weight of
    $1/\mathit{depth(l)}$, where $\mathit{depth(l)}$ is the depth of
    location $\mathit{l}$ in the AST. This means that this strategy
    assigns larger weights to locations higher in the AST.
  \item The \emph{Depth-Biased} strategy assigns to each candidate
    expression at location $\mathit{l}$ a weight of
    $\mathit{depth(l)}$, that is, larger weights are assigned to
    locations lower in the AST.
\end{itemize}

\textit{Dynamic strategies.} Dynamic strategies do not assign fixed
weights to the candidate expressions. Instead, these strategies select
an initial expression and then traverse the AST in different
directions to select another. Our tool supports the following two
dynamic strategies that differ in how they traverse the AST:
\begin{itemize}
\item The \emph{Random-Walk} strategy selects an initial candidate
  expression at the first possible location in the main function of
  the seed program. To select another expression, this strategy moves
  in a random direction in the AST, for instance, to the subsequent
  statement, the previous compound statement, or into a function call.
  \item The \emph{Guided-Walk} strategy is a variation of
    Random-Walk. In comparison, this strategy favors moves to
    locations in the AST that are likely to increase differences in
    the running times of the program analyzers, for example, by moving
    deeper in a compound statement.
\end{itemize}

%% ----------------------------------------------------------------
\subsection{Detection of Soundness and Precision Issues}
\label{subsec:IssueDetection}
%% ----------------------------------------------------------------

A common challenge for differential testing techniques is detecting
issues with a low false-positive rate, instead of reporting all found
differences. In our context, this requires determining whether the
analysis results are indeed sound or precise and, in particular,
whether any generated warnings are spurious or miss errors in the
program. To address this challenge, \tool uses two mechanisms for
detecting soundness and precision issues in the tested analyzers,
namely, the deviance and unsoundness detection mechanisms.

\textbf{Deviance detection.} Given a program variant (with a single
assertion), analyzers return one of the following verdicts:
\emph{safe} (i.e., the assertion cannot fail), \emph{unsafe} (i.e.,
the assertion can fail), or \emph{unknown} (i.e., it is unknown
whether the assertion can fail, likely because the analysis times
out).

The deviance detection mechanism is inspired by Engler et al.’s ``bugs
as deviant behavior''~\cite{EnglerChen2001}. Specifically, given a
program variant, when the majority of program analyzers return
\emph{unsafe}, \tool detects a potential soundness issue in the
deviant analyzers that return \emph{safe}. Conversely, \tool detects
a potential precision issue when a few analyzers return \emph{unsafe},
while the majority of analyzers return \emph{safe}.

We call an analyzer \emph{$\delta$-unsound} with respect to a program
variant when it returns \emph{safe} and $\delta$ other analyzers
return \emph{unsafe}. Analogously, an analyzer is
\emph{$\delta$-imprecise} with respect to a variant when it returns
\emph{unsafe} and $\delta$ other analyzers return \emph{safe}.
For each tested analyzer, \tool ranks all detected soundness
(resp. precision) issues in order of decreasing severity, where
severity is proportional to $\delta$, that is, to the number of
disagreeing analyzers.
For instance, for the program variant of \figref{fig:smack} containing
the assertion on line~\ref{line:precision}, we say that SMACK is
5-imprecise since all other analyzers disagree.

\textbf{Unsoundness detection.} Certain analyzers under-approximate
the set of possible executions in a given program. Consequently, when
such analyzers detect an error in the program, this is inevitably a
true positive (modulo bugs in the analysis itself). We consider CBMC
%% and KLEE~\cite{CadarDunbar2008}, a dynamic symbolic execution engine,
to be an under-approximating program analyzer, for instance, because
it typically analyzes the program until a bound is reached and uses
bit-precise reasoning (i.e., no over-approximation).

When such analyzers (like CBMC) find that a program variant is unsafe,
then the assertion in the variant can definitely fail. Therefore, any
other analyzer that returns \emph{safe} for the same variant is
unsound, and we call it \emph{must-unsound}. On the other hand, when
CBMC returns \emph{safe}, it is possible that it has missed an
assertion violation in the program variant due to its
under-approximation. For this reason, we do not use the results of
under-approximating analyzers to draw any definite conclusions about
imprecision in other analyzers.

%% ----------------------------------------------------------------
\section{Implementation}
\label{sec:Implementation}
%% ----------------------------------------------------------------

In this section, we present the details of our implementation, which
is open source\footnote{\anonymize{\url{https://github.com/Practical-Formal-Methods/adiff}}}.

\textbf{Check synthesis.} Recall from \algoref{alg:synthesis} that the
check-synthesis component of \tool generates a constant $\mathit{k}$,
which is used, together with a candidate expression, to form a check
for a numerical property. In general, the constants $\mathit{k}$ are
sampled randomly but with a few tweaks.

To guide the sampling, our implementation uses the following
optimizations.
First, while parsing a seed program, \tool populates a pool of
constants with any constant $\mathit{c}$ encountered in the program,
$\mathit{c + 1}$, $\mathit{c - 1}$, and boundary values (such as
$\mathit{MIN\_INT}, 0,$ and $\mathit{MAX\_INT}$). The constants in the
pool are then used to complement the randomly sampled constants.
Second, our implementation does not sample constants uniformly to
avoid frequently generating very large values. Instead, \tool
uniformly selects a bit-width for a constant and then randomly
generates a sequence of bits with this width.
Third, for each check to be synthesized, we use a type checker to
determine the type of the candidate expression and, thus, the
bit-width of the corresponding constant that will be generated (e.g.,
1 bit for expressions of type \code{bool} and 8 bits for \code{char}).

All these optimizations aim to bias or restrict the search space of
possible constants and, therefore, improve the efficiency of our tool.

\textbf{Program analysis runs.} After invoking the program analyzers,
our implementation persists the results of each analysis run in a
database. The database is extended with caching capabilities allowing
\tool to avoid re-running the analyzers on program variants that have
been previously generated.

Moreover, incorporating new program analyzers in \tool is very easy
since our implementation uses Docker containers to install, set up,
and run the analyzers. This also greatly simplifies resource
management and monitoring (e.g., memory and CPU usage, or running
time) as well as the parallelization of analysis jobs.

\textbf{Issue detection.} To detect soundness or precision issues,
users can query the database of analysis results, either by writing
their own queries or by using the default ones (described in
\secref{subsec:IssueDetection}). A query may be submitted either
through the command line or via a web-based user interface.

%% ----------------------------------------------------------------
\section{Experimental Evaluation}
\label{sec:Experiments}
%% ----------------------------------------------------------------

To evaluate the effectiveness of our approach, we apply \tool to six
state-of-the-art program analyzers. In this section, we present our
experimental setup (\secref{subsec:Setup}), give an overview of the
tested analyzers (\secref{subsec:Analyzers}), and investigate several
research questions (\secref{subsec:Results}).

%% ----------------------------------------------------------------
\subsection{Setup}
\label{subsec:Setup}
%% ----------------------------------------------------------------

We selected 1,393 seed programs (written in C) from the SV-COMP
repository of verification benchmarks~\cite{SV-COMP}. We excluded all
programs that contain ``float'', ``driver'', or ``eca-rers2012'' in
their path. The first category of excluded programs cannot be handled
by all program analyzers we tested, while the other two categories
mainly contain very large benchmarks that caused most analyzers to
reach the time limit we set for our experiments.
We also excluded any other benchmark that crashed our type checker,
for instance, some of the programs that are automatically generated by
PSYCO~\cite{GiannakopoulouRakamaric2012}.

We chose these benchmarks as seed programs because five out of the six
program analyzers we tested have participated in at least one SV-COMP
over the years. We were, therefore, confident that the analyzers would
be able to handle most of the selected programs.
Moreover, SV-COMP benchmarks typically do not exhibit arithmetic
overflows to avoid penalizing analyzers that are intentionally unsound
with respect to
overflow~\cite{LivshitsSridharan2015,ChristakisMueller2015-Clousot}.

In general, all SV-COMP benchmarks are annotated with assertions (no
other crashes should be possible), and user inputs are made explicit.
To use these benchmarks as seed programs, \tool had to preprocess
them. First, we removed all existing assertions such that the seed
programs are correct and the program analyzers do not generate any
warnings for them. Second, we ran the GCC preprocessor to eliminate
any macro usages and, thus, avoid any parsing issues in the analyzers.

Unless stated explicitly, we used the following default configuration
of \tool: a time limit of 30s, 2 CPU cores, and up to 8GB of memory
per analysis run, a budget of either 100 or 20\% of the number of
candidate expressions (whichever is smaller) per seed program, the
Uniform-Random search strategy, and a batch-check size of 4.
Recall from \secref{subsec:Synthesis} that \tool can synthesize
assertions with multiple conjuncts. These are called batch checks, and
we refer to the number of conjuncts as the `batch-check size'.

We ran our experiments on a dual hexacore
{\mbox{Intel}\textregistered~Xeon\textregistered~X5650~CPU~@~2.67GHz}
machine with 48~GiB of memory, running Debian Stretch.

%% ----------------------------------------------------------------
\subsection{Program Analyzers}
\label{subsec:Analyzers}
%% ----------------------------------------------------------------

We selected the analyzers under test such that they cover a wide range
of different analysis techniques. In addition, we only chose mature
tools that are under active development. We give a short description
of each analyzer below. Note that, unless otherwise stated, we used
their default configuration.

\textbf{CBMC.}
CBMC~\cite{ClarkeKroening2004} is a bit-precise bounded model checker
that unrolls loops and expresses verification conditions as SMT
queries over bit-vectors. We used version \texttt{\small{5.3}} of the
tool.

\textbf{CPAchecker.}
CPAchecker~\cite{BeyerKeremoglu2011} is a software model checker that
incorporates different program-analysis techniques, such as predicate
abstraction~\cite{GrafSaidi1997,BallMajumdar2001}, lazy
abstraction~\cite{HenzingerJhala2002}, and
k-induction~\cite{DonaldsonHaller2011}. We used the development
version \texttt{\small{51.7-svn28636}} of the tool, which incorporates fixes for two
soundness issues that \tool detected.
Note that CPAchecker won the first place in SV-COMP'18.

\textbf{Crab.}
Crab~\cite{GangeNavas2016-Domain,GangeNavas2016-Sparsity} is an
abstract interpreter that supports several abstract
domains~\cite{CousotCousot1977,CousotCousot1979}. Its default
configuration uses the Zones domain~\cite{Mine2004}, and we enabled
inter-procedural analysis. The tool was built from commit
\texttt{\small{5dd7a00b5b}}.

%% \textbf{KLEE}
%% %
%% KLEE~\cite{CadarDunbar2008} is a dynamic symbolic
%% execution~\cite{GodefroidKlarlund2005,CadarEngler2005} tool (also known as concolic
%% testing). We configured KLEE to use depth-first search and limited forks to 64. Programs
%% were compiled to LLVM bytecode using Clang without optimizations. We followed the
%% experimental build instructions from the KLEE
%% website\footnote{\url{http://klee.github.io/build-llvm38/}} to use it with a more recent
%% version of LLVM (3.8).

\textbf{SeaHorn.}
SeaHorn~\cite{GurfinkelKahsai2015} is a software model checker that
expresses verification conditions as Horn-clauses and uses existing
solvers to discharge them. Its default configuration uses
Spacer~\cite{KomuravelliGurfinkel2013,KomuravelliGurfinkel2014}, which
is a fork of Z3~\cite{deMouraBjorner2008} with a variant of the
IC3/PDR~\cite{Bradley2011} model-checking algorithm for solving
verification conditions. The tool was built from commit
\texttt{\small{59c4a917a595}}.

\textbf{SMACK.}
SMACK~\cite{RakamaricEmmi2014} is software model checker that
translates C programs to Boogie~\cite{BarnettChang2005}, which can be
checked by a number of different verification back-ends. We used
version \texttt{\small{1.9}} with the default configuration, which
runs the Corral verifier~\cite{LalQadeer2012}. We also enabled the
\texttt{svcomp} extension and set the loop-unrolling bound to 1,000.
Note that SMACK won the second place in SV-COMP'17.

\textbf{Ultimate Automizer.}
Ultimate Automizer (or UAutomizer) is a software model checker that
uses an automata-based verification
approach~\cite{HeizmannHoenicke2013,DietschHeizmann2017,HeizmannChen2018}. We used version
\texttt{\small{0.1.23}}.
Note that UAutomizer won the second place in SV-COMP'18.

For all program analyzers that support this, we set the machine
architecture to 32-bit. We also provided a default LTL-specification
file to any analyzer that requires an explicit reachability property
for checking assertions.

%% ----------------------------------------------------------------
\subsection{Results}
\label{subsec:Results}
%% ----------------------------------------------------------------

We break our experimental results down into five categories, each
investigating a different research question.\\

\begin{figure*}[t]
\centering
\scalebox{0.85}{
  \begin{tikzpicture}
    \begin{axis}[
      % ybar stacked,
      ybar ,
      xticklabels={\textbf{CBMC},\textbf{CPAchecker},\textbf{Crab},\textbf{SeaHorn},\textbf{SMACK},\textbf{UAutomatizer}},
      xtick=data,
      ylabel={\textbf{Number of soundness issues}},
      legend cell align=left,
      axis y line*=none,
      axis x line*=bottom,
      width=18cm,
      height=4.5cm,
      ymin=0,
      ymax=65,
      bar width=0.45cm,
      area legend,
      nodes near coords,
      % every node near coord/.append style={rotate=90, anchor=west}
      ]

      \pgfplotstableread{data/delta-unsoundness.dat}\unsoundness

      \addplot+[ybar, Purple, pattern color=Purple!70, pattern=north east lines]
          table[x expr=\coordindex, y=cbmc] from \unsoundness;
      \addplot+[ybar, Red]
          table[x expr=\coordindex, y=x5] from \unsoundness;
      \addplot+[ybar, Crimson, pattern color=Crimson, pattern=crosshatch dots]
          table[x expr=\coordindex, y=x4] from \unsoundness;
      \addplot+[ybar, DarkOrange, pattern color=DarkOrange!70, pattern=north west lines]
          table[x expr=\coordindex, y=x3] from \unsoundness;
      \legend{must, $\delta=5$, $\delta=4$, $\delta=3$}
    \end{axis}
  \end{tikzpicture}
}
\caption{Soundness issues detected for each program analyzer.}
\label{fig:soundnessBarChart}
\vspace{-1.5em}
\end{figure*}
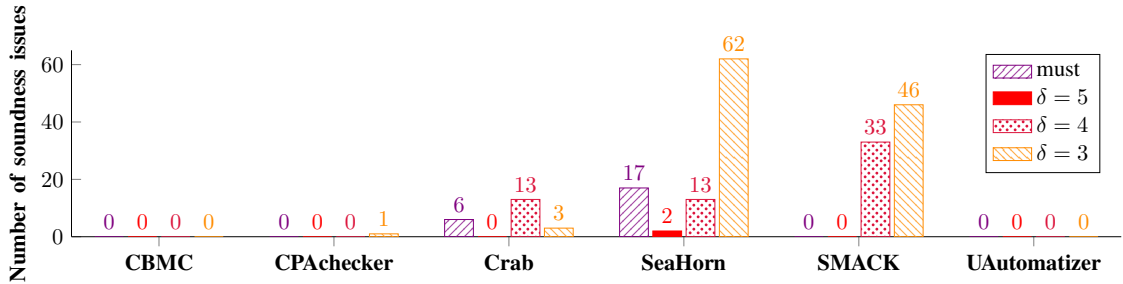

\textbf{RQ1: Does $\bm{\alpha}$-Diff find soundness and precision issues?}
Given as input the 1,393 seed programs, \tool generated 25,960 program
variants. \figref{fig:soundnessBarChart} shows the number of potential
soundness issues that \tool detected in the tested program analyzers
when running them on the generated variants. Each soundness issue
corresponds to a program variant that revealed an analyzer to be must-
or $\delta$-unsound. As shown in the figure, our technique detected a
significant number of potential issues in four program analyzers.
Note that some of these issues might expose the same source of
unsoundness in an analyzer. For example, assume that \tool generates
several variants that use bitwise arithmetic in their assertions. For
each of these variants, our tool could potentially report a soundness
issue in any analyzer that does not support bit-precise reasoning.

% \begin{figure}[t]
% \centering
% \scalebox{0.85} {
%   % barchart with klee/cbmc soundness
%   \begin{tikzpicture}
%     \begin{axis}[
%       ybar,
%       xticklabels from table={data/klee-cbmc-unsoundness.dat}{verifier},
%       xtick=data,
%       xticklabel style={rotate=-45},
%       ylabel={number of issues},
%       legend cell align=left,
%       legend style={at={(0.05,0.8)}, anchor=west},
%       axis y line*=none,
%       axis x line*=bottom,
%       width=10cm,
%       height=6cm,
%       ymin=0,
%       ymax=45,
%       bar width=0.45cm,
%       area legend,
%       ]
%       \pgfplotstableread{data/klee-cbmc-unsoundness.dat}\tbl;
%       \addplot+[ybar, Crimson, pattern color=Crimson, pattern=crosshatch dots]
%         table[x expr=\coordindex, y=klee] from \tbl;
%       \addplot+[ybar, DarkOrange, pattern color=DarkOrange, pattern=north east lines]
%         table[x expr=\coordindex, y=cbmc] from \tbl;
%       \legend{klee-unsound, cbmc-unsound}

%       \end{axis}
%     \end{tikzpicture}
% }
% \caption{KLEE/CBMC-soundness issues found for each analyzer. \done{semi-final}}
% \label{fig:KLEECBMCsoundnessBarChart}
% \end{figure}

We manually inspected all detected issues from
\figref{fig:soundnessBarChart} and reported unique sources of
unsoundness to the designers of Crab, SeaHorn, and SMACK.
Note, however, that \tool had previously found two soundness issues in
CPAchecker, which were reported to the tool designers early on and
were
fixed\footnote{\anonymize{\url{https://groups.google.com/d/msg/cpachecker-users/3JCOeNuoleA/fpr8ElaaBgAJ}}}\footnote{\anonymize{\url{https://groups.google.com/d/msg/cpachecker-users/3JCOeNuoleA/YDT7LokZBwAJ}}}. We
used the patched version of CPAchecker for our experiments.
We discuss the reaction of all designers in RQ2.

As shown in the figure, CPAchecker is found to be 3-unsound for one
program variant. This issue is actually a source of imprecision in
three other analyzers, namely, in Crab, SeaHorn, and SMACK, and is
related to bit-precise reasoning. In general, we observed that the
false-positive rate of our technique depends on two main factors.
First, if some of the tested program analyzers are imprecise for a
given variant, \tool detects soundness issues in the remaining
analyzers, which are, however, sound. Second, any issues that are
reported for $\delta$-unsound analyzers, where $\delta$ is small, are
likely false positives. For example, when a program analyzer is
1-unsound, there exists only one disagreeing analyzer. In fact,
inspecting issues found for $\delta$-unsound analyzers, where
$\delta$ is large, significantly reduces the number of false positives
caused by imprecision in some of the other analyzers.
In the results of \figref{fig:soundnessBarChart}, we did not find any
false positives when $\delta \geq 4$ or when inspecting the issues
that were detected for must-unsound analyzers.

% Notes about our 'false positives'

% UNSOUNDNESS
%  cbmc(d=2) : bitvector/interleave_bits.... imprecision in crab and seahorn
%  - e.g. (x | xx << 1u) != 1u   is tautological, but crab & seahorn cannot  prove it.
%  cpachecker(d=3) : c/bitvector/sum02_true-unreach-call_true-no-overflow.i
%   - tough problem: Proving that 86014472 is not a Gaussian sum, e.g. not
%   exists an n such that n * (n + 1) / 2 = 86014472. (imprecision in crab, seahorn, smack - same as above)
%  seahorn(d=5): c/reducercommuntativity/max05/max10...
%   - asserting value of uninitialized array, seahorn: 'unsat'
%  seahorn(d=4): see above
%  smack (d=4):
%   - loops/vogal_true : uninitialized variable
%   - array-examples/standard_inti2... : 'hacky heuristic'
%

\begin{figure}[b!]
\vspace{-2em}
\centering
\scalebox{0.85}{
  \begin{tikzpicture}
    \pgfplotstableread{data/delta-incompleteness.dat}\incompleteness
    \begin{axis}[
      ybar,
      xticklabels={\textbf{CBMC},\textbf{CPAchecker},\textbf{Crab},\textbf{SeaHorn},\textbf{SMACK},\textbf{UAutomatizer}},
      xtick=data,
      xticklabel style={rotate=-45},
      ylabel={\textbf{Number of precision issues}},
      legend cell align=left,
      legend style={at={(0.05,0.8)}, anchor=west},
      axis y line*=none,
      axis x line*=bottom,
      width=10cm,
      height=4.2cm,
      ymin=0,
      ymax=35,
      restrict y to domain*=0:40,
      visualization depends on=rawy\as\rawy,
      after end axis/.code={ % Draw line indicating break
                  \draw [ultra thick, white, decoration={snake, amplitude=1pt}, decorate] (rel axis cs:0,1.05) -- (rel axis cs:1,1.05);
              },
      clip=false,
      bar width=0.30cm,
      area legend,
      nodes near coords={\pgfmathprintnumber{\rawy}},
      every node near coord/.append style={rotate=90, anchor=west}
      ]

      \addplot+[ybar, Red]
        table[x expr=\coordindex, y=x5] from \incompleteness;
      \addplot+[ybar, Crimson, pattern color=Crimson, pattern=crosshatch dots]
        table[x expr=\coordindex, y=x4] from \incompleteness;
      \addplot+[ybar, DarkOrange, pattern color=DarkOrange, pattern=north west lines]
        table[x expr=\coordindex, y=x3] from \incompleteness;

      \legend{$\delta=5$, $\delta=4$, $\delta=3$}
      \end{axis}
  \end{tikzpicture}
}
\caption{Precision issues detected for each analyzer.}
\label{fig:imprecisionBarChart}
\end{figure}
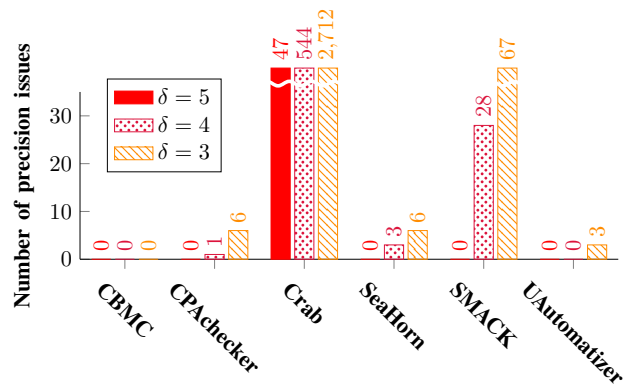

\figref{fig:imprecisionBarChart} shows the number of precision
issues that \tool found for the same program variants.
Although the number of issues is significant, the majority of these do
not correspond to bugs in the analyzers. Instead, most of the
precision issues are either intended by the analysis designers (for
instance, imprecise reasoning about numeric types) or inherent to the
analysis (for example, imprecision in non-relational abstract domains,
such as Intervals).

Overall, \tool found many more precision issues in Crab in comparison
to the other analyzers. These issues, however, are intended since Crab
favors performance over precision, similarly to several other abstract
interpreters. Manual inspection of a selection of these issues showed
that the program variants for which Crab is imprecise exhibited at
least one of the following features: (1)~usage of pointers,
(2)~bitwise operations, (3)~invariants expressing parity of variables.

%% \begin{table}[t]
%% \centering
%% \scalebox{1}{
%%   \begin{tabular}{|l|S[table-format=1.2,table-comparator=true]|S[table-format=1.2,table-comparator=true]|S[table-format=1.2]|S[table-format=1.2]|S[table-format=1.2,table-comparator=true]|}
%%     \hline
%%     \diagbox{$\bm{D_i}$}{$\bm{D_j}$} & \multicolumn{1}{c|}{\textbf{int}}  & \multicolumn{1}{c|}{\textbf{oct}}  & \multicolumn{1}{c|}{\textbf{pk}}   & \multicolumn{1}{c|}{\textbf{rtz}}  & \multicolumn{1}{c|}{\textbf{zones}}\\
%%     \hline
%%     \textbf{int}                    & 1.00 & 0.97 & 0.97 & 0.97 & 0.89 \\
%%     \textbf{oct}                    & 0.99 & 1.00 & 0.99 & 0.99 & 0.99 \\
%%     \textbf{pk}                    & 0.99 & 0.99 & 1.00 & 0.98 & 0.99 \\
%%     \textbf{rtz}                    & < 1.00  &  < 1.00 & 0.99 & 1.00 & < 1.00 \\
%%     \textbf{zones}                  & 1.00 & < 1.00 & 0.99 & 0.99 & 1.00\\
%%     \hline
%%   \end{tabular}
%% }
%% \caption{Relative precision for abstract domains of Crab.}
%% \label{tab:CrabDomains}
%% \end{table}

\begin{table*}[t]
\centering
\scalebox{0.85}{
  \begin{tabular}{|p{1.6cm}|S[table-format=1.2]|S[table-format=1.2]|S[table-format=1.2,table-comparator=true]|S[table-format=1.2]|S[table-format=1.2]|S[table-format=1.2,table-comparator=true]|}
    \hline
      \diagbox[innerwidth=1.6cm]{$\bm{A_i}$}{$\bm{A_j}$} & \multicolumn{1}{c|}{\textbf{CBMC}} & \multicolumn{1}{c|}{\textbf{CPAchecker}} & \multicolumn{1}{c|}{\textbf{Crab}} & \multicolumn{1}{c|}{\textbf{SeaHorn}} & \multicolumn{1}{c|}{\textbf{SMACK}} & \multicolumn{1}{c|}{\textbf{UAutomatizer}}\\
      \hline

      \textbf{CBMC}                                     & 1.00 & 0.01       & 0.01      & 0.01    & 0.01  & 0.01 \\
      \textbf{CPAchecker}                               & 0.71 & 1.00       & 0.99      & 0.95    & 0.97 & 0.99 \\
      \textbf{Crab}                                & 0.37 & 0.63       & 1.00      & 0.62    & 0.88 & 0.68 \\
      \textbf{SeaHorn}                                  & 0.77 & 0.98       & <1.00      & 1.00    & 0.99  & < 1.00 \\
      \textbf{SMACK}                                    & 0.72 & 0.53       & 0.76      & 0.53    & 1.00  & 0.61 \\
      \textbf{UAutomatizer}                               & 0.57 & 0.82       & 0.89      & 0.81    & 0.93  & 1.00\\
      \hline
    \end{tabular}
  }
  \caption{Relative precision for the tested program analyzers.}
  \label{tab:RelPrecision}
  \vspace{-1.5em}
\end{table*}

In addition to detecting issues in program analyzers, \tool can also
compare their relative soundness and precision. To determine the
relative precision (resp. soundness) of an analyzer $A_i$ with respect
to another analyzer $A_j$, our technique computes the probability that
$A_i$ returns \emph{safe} (resp. \emph{unsafe}) given that $A_j$ returns \emph{safe}
(resp. \emph{unsafe}). \tabref{tab:RelPrecision} shows these probabilities
for all analyzers under test.
Note that $<$1.00 stands for a probability between 0.99 and 1.00.

From the first row of the table, we observe that CBMC verifies only
1\% of the variants that all other analyzers prove safe. This is due
to the fact that CBMC uses bounded model checking, which might fail to
explore all program paths within a certain time limit.
In contrast, CPAchecker verifies 71\% of the variants that CBMC proves
safe.
As another example, SeaHorn verifies almost all variants that
UAutomizer proves safe, while UAutomizer verifies only 81\% of the
variants that SeaHorn proves safe. This clearly indicates that SeaHorn
is more precise on the generated program variants.
On the other hand, \tool did not detect any soundness issues for
UAutomatizer (see \figref{fig:soundnessBarChart}), which could
potentially explain the higher precision of SeaHorn on some variants.
In general, analyzers might gain in precision when sacrificing
soundness since they consider fewer program executions.

\tabref{tab:SmackVsCrab} shows a direct comparison between SMACK and
Crab, which implement very different analysis techniques. Note that
$>$0\% stands for a percentage between 0 and 1. As shown in the table,
for 3\% of the program variants, Crab reports unsafe whereas SMACK
returns safe. Both tools prove 23\% of the variants safe and never
return unknown for the same program variant.
This suggests that these analysis techniques have complementary
strengths and weaknesses.

\begin{table}[t]
\centering
\scalebox{0.85}{
  \begin{tabular}{|l|S[table-format=2.0,table-space-text-post=\si{\percent}]|S[table-format=2.0,table-space-text-post=\si{\percent}]|S[table-format=2.0,table-comparator=true,table-space-text-post=\si{\percent}]|}
    \hline
    \diagbox{\textbf{SMACK}}{\textbf{Crab}} & \multicolumn{1}{c|}{\textbf{unsafe}} & \multicolumn{1}{c|}{\textbf{unknown}} & \multicolumn{1}{c|}{\textbf{safe}}\\
    \hline
    \textbf{unsafe}                & 14\si{\percent}   & 22\si{\percent}    & > 0\si{\percent}  \\  %%& 36\%  \\
    \textbf{unknown}               & 30\si{\percent}  & 0\si{\percent}     & 7\si{\percent}      \\  %%& 37\% \\
    \textbf{safe}                  & 3\si{\percent}    & 0\si{\percent}     & 23\si{\percent}     \\  %%& 26\% \\
    \hline
%%    Totals                & 47\%   & 22\%    & 30\%  & 100\%
  \end{tabular}
}
\caption{Comparison of results for SMACK and Crab.}
\label{tab:SmackVsCrab}
  \vspace{-1.5em}
\end{table}

In addition to comparing the results of different analyzers, \tool can
also be used to compare different configurations of the same analyzer.
\tabref{tab:CrabOctVsPoly} shows a direct comparison of two Crab
configurations, each using a different abstract domain, namely,
Octagons~\cite{Mine2006} and Polyhedra~\cite{CousotHalbwachs1978}.  As
shown in the table, there is a small number of program variants that
are verified with the Octagons domain but not with Polyhedra, although
in theory Polyhedra is strictly more precise than Octagons.
As pointed out by the designer of
Crab\footnote{\anonymize{\url{https://github.com/seahorn/crab-llvm/issues/18}}},
this mismatch is due to the fact that the domains use different
widening operations~\cite{CousotCousot1976,CousotCousot1992} to speed
up convergence of the fixed-point computation. This is a known
caveat~\cite{MonniauxLeGuen2012} and was independently evaluated in a
recent paper comparing different abstract
domains~\cite{AmatoRubino2018}.

In \tabref{tab:CrabDomains}, we use \tool to compare the relative
precision of several abstract domains of Crab, namely, Intervals,
Octagons, Polyhedra, RTZ (i.e., the reduced product of disjunctive
Intervals and Zones), and Zones.
Across the domains, the differences in precision are small for the
generated variants. However, not surprisingly, the Intervals domain is
typically less precise than the others. For instance, Intervals can
only verify 89\% of the variants that are proved safe with Zones. On
the other hand, the very precise Polyhedra domain can only verify 99\%
of the variants that are proved safe with Intervals. As previously
explained, this is due to the widening operation.

\textbf{RQ2: Are the issues relevant for designers of analyzers?}\\
\noindent
To determine whether the issues that \tool reports are relevant to
analysis designers, we inspected all detected soundness issues as well
as a selection of precision issues.  Overall, we found ten unique
soundness and precision issues in four (out of six) program analyzers
(excluding CBMC and UAutomizer). We reported nine of these
issues to the designers of the four analyzers.  All reported issues
were confirmed and three (in CPAchecker and Crab) were fixed in only a
few hours each. We now discuss the detected issues in detail.

\emph{CPAchecker.}
Although for the experiments in this paper we used the patched version
of CPAchecker, \tool detected two sources of unsoundness in this
analyzer, which were immediately fixed.
The first soundness
issue\footnote{\anonymize{\url{https://groups.google.com/d/msg/cpachecker-users/3JCOeNuoleA/fpr8ElaaBgAJ}}}
was revealed in the simplified version of a program variant shown in
\figref{fig:CPAchecker} (line~\ref{line:CPAchecker2} should be
ignored).
In this program, \code{x} is assigned a non-deterministic integer and,
therefore, the assertion on line~\ref{line:CPAchecker1} can fail. A
previous version of CPAchecker missed this assertion violation due to
a bug in its invariant-generation component, which was unsound when
trying to obtain information about the factors of a multiplication
whose product was zero.

\begin{table}[t]
\centering
\scalebox{0.85}{
  \begin{tabular}{|l|S[table-format=2.0,table-comparator=true,table-space-text-post=\si{\percent}]|S[table-format=2.0,table-comparator=true,table-space-text-post=\si{\percent}]|S[table-format=2.0,table-comparator=true,table-space-text-post=\si{\percent}]|}
    \hline
    \diagbox{\textbf{oct}}{\textbf{pk}} & \multicolumn{1}{c|}{\textbf{unsafe}} & \multicolumn{1}{c|}{\textbf{unknown}} & \multicolumn{1}{c|}{\textbf{safe}}\\
    \hline
    \textbf{unsafe} & 43\si{\percent} & 1\si{\percent} & > 0\si{\percent} \\
    \textbf{unknown} & > 0\si{\percent} & 10\si{\percent} & > 0\si{\percent} \\
    \textbf{safe} & > 0\si{\percent} & > 0\si{\percent} & 45\si{\percent} \\
    \hline
  \end{tabular}
}
\caption{Comparison of Crab Octagons and Polyhedra.}
\label{tab:CrabOctVsPoly}
  \vspace{-1.5em}
\end{table}

The second
issue\footnote{\anonymize{\url{https://groups.google.com/d/msg/cpachecker-users/3JCOeNuoleA/YDT7LokZBwAJ}}}
was revealed in the simplified variant of \figref{fig:CPAchecker} when
considering line~\ref{line:CPAchecker2} (and ignoring
line~\ref{line:CPAchecker1}).
According to the C standard, the expression \code{x == 1} evaluates to
an integer of value \code{0} or \code{1}, which is never equal to
\code{99}. Consequently, the assertion on line~\ref{line:CPAchecker2}
can fail since \code{x} is assigned a non-deterministic
integer. CPAchecker missed this assertion violation due to a bug in
its value analysis, which was unsound when analyzing nested binary
expressions such as the property asserted above.

% CPAchecker bug fixes:
% - Fix a bug in invariant generation where the reverse computation of a multiplication to
% obtain information about its factors was unsound if the multiplication result was zero
% (https://gitlab.com/sosy-lab/software/cpachecker/commit/aa3b3fbd15f217a519f6280ba1543c5314915134)
% - Fix a bug where value analysis was unsound when evaluating assumptions over nested
% binary expressions such as '(x == c) == d', because it would not only 'learn' from the
% outer expression, but would assume the truth of inner expressions as well, which is
% obviously wrong
% (https://gitlab.com/sosy-lab/software/cpachecker/commit/8d7645f938f6ca08dcae447f10f4ab99d12f9225)

% Crab bug fix:
% - [FIXED] detection of recursive funtions in the inter-procedural analysis
% (https://github.com/seahorn/crab/commit/11fffc78ff7851fa950953a9fd31c980dece8a79)

% Crab (not fixed): undefined behavior (https://github.com/seahorn/crab-llvm/issues/20)

\emph{Crab.}
Our technique detected two soundness issues in Crab. The first
issue\footnote{\anonymize{\url{https://github.com/seahorn/crab-llvm/issues/20}}}
made the inter-procedural analysis of Crab unsound in the presence of
recursion, and the bug was immediately fixed.

The second issue (which was reported together with the first) is
caused by Crab's LLVM-based~\cite{LattnerAdve2004} front-end, which
may optimize the program by exploiting undefined behavior.
For example, several seed programs contain uninitialized variables.
According to the C standard, the behavior of a program that reads from
such variables is undefined, that is, any behavior is correct.
In these cases, a compiler pass may under-approximate the behavior of
the program, for instance, by assuming that any read from an
uninitialized variable returns \code{0}, to optimize the executable
code. However, this under-approximation potentially leads to
unsoundness in program analyzers that analyze the optimized code,
given that the original program may fail when compiled without this
optimization or with a different compiler.

We also reported two imprecision
issues\footnote{\anonymize{\url{https://github.com/seahorn/crab-llvm/issues/18}}}
in the Polyhedra and Octagons domains of Crab. In particular, the less
precise Intervals domain was able to verify the assertion in a program
variant, whereas the more precise Polyhedra domain found the variant
unsafe. As discussed in RQ1, such precision issues are possible for
abstract domains with a widening
operation.
The issue was similar for the Octagons domain, which typically ignores
dis-equalities. In contrast, given the interval $x = [0, 10]$ and
the constraint $x \neq 10$, the Intervals domain does compute the more
precise interval $x = [0, 9]$.

\begin{table}[t]
\centering
\scalebox{0.85}{
  \begin{tabular}{|l|S[table-format=1.2,table-comparator=true]|S[table-format=1.2,table-comparator=true]|S[table-format=1.2]|S[table-format=1.2]|S[table-format=1.2,table-comparator=true]|}
    \hline
    \diagbox{$\bm{D_i}$}{$\bm{D_j}$} & \multicolumn{1}{c|}{\textbf{int}}  & \multicolumn{1}{c|}{\textbf{oct}}  & \multicolumn{1}{c|}{\textbf{pk}}   & \multicolumn{1}{c|}{\textbf{rtz}}  & \multicolumn{1}{c|}{\textbf{zones}}\\
    \hline
    \textbf{int}                    & 1.00 & 0.97 & 0.97 & 0.97 & 0.89 \\
    \textbf{oct}                    & 0.99 & 1.00 & 0.99 & 0.99 & 0.99 \\
    \textbf{pk}                    & 0.99 & 0.99 & 1.00 & 0.98 & 0.99 \\
    \textbf{rtz}                    & < 1.00  &  < 1.00 & 0.99 & 1.00 & < 1.00 \\
    \textbf{zones}                  & 1.00 & < 1.00 & 0.99 & 0.99 & 1.00\\
    \hline
  \end{tabular}
}
\caption{Relative precision for abstract domains of Crab.}
\label{tab:CrabDomains}
\vspace{-1.5em}
\end{table}

\emph{SeaHorn.}
We reported a
soundness\footnote{\anonymize{\url{https://github.com/seahorn/seahorn/issues/152}}}
and a precision
issue\footnote{\anonymize{\url{https://github.com/seahorn/seahorn/issues/157}}}
to the designers of SeaHorn, who confirmed both issues.
The soundness issue was caused by SeaHorn's LLVM-based front-end,
which is slightly different than Crab's and, thus, results in
different unsound results.

Regarding the precision issue, the designers of SeaHorn explained that
it is due to the conservative handling of bitwise operations and
numeric types. In particular, all numeric types are abstracted into
arbitrary-precision signed integers.

\emph{SMACK.}
We reported the soundness and precision
issues\footnote{\anonymize{\url{https://github.com/smackers/smack/issues/324}}}
that are caused by the size-reduction heuristic in SMACK (see
\secref{sec:Overview} for details). Although confirmed, these issues
were not fixed since this behavior was intended by the designers.

We also found several other soundness issues, which were due to
optimizations by SMACK's LLVM-based front-end, just like in Crab and
SeaHorn. We did not report these issues to the designers since their
cause is clear.

In general, the reaction of the analysis designers to all reported
issues shows that \tool can detect important sources of unsoundness
and imprecision. This is especially the case since the tested
analyzers are mature tools that are under active
development. Moreover, five of these analyzers (excluding Crab) have
participated in SV-COMP, which did not reveal any of the above bugs.

\textbf{RQ3: Are the controversial issues relevant for users?}
Half of the tested program analyzers, namely Crab, SeaHorn, and SMACK,
might be unsound in the presence of undefined behavior. As discussed
earlier, this unsoundness is caused by compiler optimizations that
under-approximate the behavior of the program. We call such soundness
issues controversial because different compilers are inconsistent in
reasoning about undefined behavior and, consequently, the results of
analyzers that analyze executable code can be contradictory.

To shed more light on what users expect from program analyzers in the
presence of undefined behavior, we performed a survey of 16
professional developers, who we hired on
Upwork\footnote{\tiny\url{https://www.upwork.com/}}.
To screen the candidates, we used two short interview questions (about
type-conversion rules and pointer usage in C). Out of the candidates
that replied correctly, we selected those that had experience with C.

The survey contained 9 short tasks. Each task included a short C
program, which was a simplified version of a program variant generated
by \tool. For every task, we asked whether the assertion in the given
program can fail, and just like a program analyzer, a survey
participant could respond with \emph{yes} (i.e., unsafe), \emph{no}
(i.e., safe), and \emph{I don't know} (i.e., unknown).

\begin{figure}[t]
\begin{lstlisting}[style=clang]
int main() {
  int x = ?\nd?;
  °assert  2 * x != 0;° ?\label{line:CPAchecker1}?
  °assert (x == 1) != 99 && x == 1;° ?\label{line:CPAchecker2}?
  return x == 1;
}
\end{lstlisting}
\vspace{-0.8em}
\caption{Soundness issues in CPAchecker.}
\label{fig:CPAchecker}
\vspace{-1.2em}
\end{figure}

To pilot the survey tasks, we sent the survey to 4 students and
interns who study Computer Science and already have a Bachelor's
degree. We asked these participants if they found any portion of the
survey difficult to understand and requested their feedback. Their
responses were solely used to improve the survey.

After finalizing the tasks, we sent the survey to the professional
developers. The tasks were presented to the developers in a randomized
order, but in total, the survey included 6 non-controversial and 3
controversial questions.
The non-controversial questions asked about programs for which most
analyzers were in agreement regarding their safety, whereas the
controversial questions asked about unsafe programs that contained
undefined behavior.
We used the non-controversial tasks to exclude participants who gave
the wrong answer to at least 4 (out of 6) of these questions. Based on
this threshold, we excluded 4 (out of 16) survey participants.

\tabref{tab:SurveyPros} shows the survey responses from the 12
developers that we did not exclude. The first column shows the task
identifier: tasks 1--6 are non-controversial and tasks 7--9 are
controversial. Next to each task identifier, we indicate whether there
exists an execution of the corresponding program that fails. For
example, when the executable code of the programs in tasks 7--9 is not
optimized, the assertions can be violated. The remaining columns of
the table show the survey responses categorized as unsafe,
unknown, and safe.

\begin{table}[b]
\centering
\begin{tabular}{|S[table-format=1.0]l|S[table-format=2.0]|S[table-format=1.0]|S[table-format=2.0]|}
\hline
\multicolumn{2}{|c|}{\textbf{Task}} & \multicolumn{3}{c|}{\textbf{Survey Response}}\\
\multicolumn{2}{|c|}{\textbf{Identifier}} & \multicolumn{1}{c|}{\textbf{unsafe}} & \multicolumn{1}{c|}{\textbf{unknown}} & \multicolumn{1}{c|}{\textbf{safe}}\\
\hline
  1 & (unsafe)  & \maxhl{11}  & 0    & 1\\
  2 & (unsafe)  & \maxhl{12}  & 0    & 0\\
  3 & (unsafe)  & \maxhl{11}  & 0    & 1\\
%%\hline
  4 & (safe)    & 2   & 0    & \maxhl{10}\\
  5 & (safe)    & 0   & 0    & \maxhl{12}\\
  6 & (safe)    & 0   & 0    & \maxhl{12}\\
\hline
  7 & (unsafe)  & \maxhl{10}  & 1    & 1\\
  8 & (unsafe)  & \maxhl{10}  & 1    & 1\\
  9 & (unsafe)  & \maxhl{10}  & 1    & 1\\
\hline
\end{tabular}
\caption{Survey responses from professional C developers.}
\label{tab:SurveyPros}
\vspace{-2.5em}
\end{table}

As shown in the table, the majority of the participants (10 out of 12
professional developers) considered the controversial programs to be
unsafe. This clearly suggests that program analyzers should treat
undefined behavior as non-determinism, instead of optimizing it away.
%% This result could be an important take away for analysis designers.
On the other hand, the 4 excluded developers were not able to give correct
answers to at least 4 questions from tasks 1--6, and the remaining 12
developers gave 4 wrong answers to these tasks (see
\tabref{tab:SurveyPros}). This is a strong indication that even
professional C developers benefit from program analysis.

%% We also sent the same survey to researchers and students in the
%% authors' organization. We received 13 complete responses that were
%% very similar to the responses from the professional developers. We
%% will make all survey results available.

\textbf{{RQ4}: What is the effect of the search strategy?}
To generate a seed-program variant, our technique explores the search
space of all possible candidate expressions using five different
search strategies (see \secref{subsec:Synthesis}).
To evaluate how each search strategy affects the number of detected
issues, we ran \tool on 20 seed programs, which were randomly selected
from the seed programs used in the evaluation of RQ1. For each run of
\tool on the seed programs, we enabled a different search strategy,
and we measured the number of soundness and precision issues that were
found.
We used the default configuration of our tool, but with a batch-check
size of 1, to prevent batch checks from influencing the results.

\tabref{tab:SearchStrategies} shows the effect of the five search
strategies on the number of detected issues. The first column of the
table shows the search strategy, the second the cumulative number of
soundness issues detected in all must-unsound analyzers, the third the
soundness issues detected in ${\geq 3}$-unsound analyzers (that is, in
$3$-, $4$-, and $5$-unsound analyzers), and the fourth the precision
issues detected in ${\geq 3}$-imprecise analyzers.
In general, the results suggest that the static search strategies are
more effective in detecting soundness and precision issues than the
dynamic strategies. Among the static strategies, the Uniform-Random
strategy helps find the most soundness issues, although the
differences are small. Among the dynamic strategies, the Random-Walk
strategy performs the worst.
We also observed that Breadth-Biased and Guided-Walk each detect a soundness
issue that is not found by any other strategy.

\begin{table}[t]
\centering
\scalebox{0.85}{
  \begin{tabular}{|l|S[table-format=1.0]|S[table-format=1.0]|S[table-format=2.0]|}
    \hline
    \multicolumn{1}{|c|}{\textbf{Search Strategy}} & \multicolumn{3}{c|}{\textbf{Number of Detected Issues}}\\
    & \multicolumn{1}{c|}{\textbf{must-unsound}} & \multicolumn{1}{c|}{\textbf{$\bm{\geq 3}$-unsound}} & \multicolumn{1}{c|}{\textbf{$\bm{\geq 3}$-imprecise}}\\
    \hline
    Uniform-Random & \maxhl{2}    & \maxhl{2}       & 71 \\
    Breadth-Biased & 1            & \maxhl{2}       & 70 \\
    Depth-Biased   & 1            & 1               & \maxhl{77} \\
    Random-Walk    & 0            & 0               & 59 \\
    Guided-Walk    & 1            & \maxhl{2}       & 39 \\
    \hline
  \end{tabular}
}
\caption{Effect of search strategies on the number of detected issues.}
\label{tab:SearchStrategies}
\vspace{-1.5em}
% for cbmc-unsound:
% - smart: ldv-memsafety/ArraysWithLenghtAtDeclaration_false-valid-deref-write.c
% - bfs  : ldv-memsafety/ArraysWithLenghtAtDeclaration_false-valid-deref-write.c
% - dfs  : ldv-memsafety/ArraysWithLenghtAtDeclaration_false-valid-deref-write.c
% - uni  : ldv-memsafety/ArraysWithLenghtAtDeclaration_false-valid-deref-write.c
% - walk : none

% for >=3-unsound:
% - bfs:
%   + ldv-memsafety/ArraysWithLenghtAtDeclaration_false-valid-deref-write.c
%   + c/array-tiling/mbpr4_true-unreach-call.i
% - dfs:
%   + ldv-memsafety/ArraysWithLenghtAtDeclaration_false-valid-deref-write.c
% - smart:
%   + ldv-memsafety/ArraysWithLenghtAtDeclaration_false-valid-deref-write.c
%   + loop-new/count_by_1_true-unreach-call_true-termination.i)
% - uniform:
%   + ldv-memsaftey/ArraysWithLenghtAtDeclaration_false-valid-deref-write.c
% - random-walk: none
\end{table}

\begin{table}[b]
\vspace{-.5em}
\centering \scalebox{0.85}{
  \begin{tabular}{|S[table-format=2.0]|S[table-format=2.0]|S[table-format=2.0]|S[table-format=3.0]|}
    \hline
      \multicolumn{1}{|c|}{\textbf{Batch-Check Size}} & \multicolumn{3}{c|}{\textbf{Number of Detected Issues}}\\
      & \multicolumn{1}{c|}{\textbf{must-unsound}} & \multicolumn{1}{c|}{\textbf{$\bm{\geq 3}$-unsound}} & \multicolumn{1}{c|}{\textbf{$\bm{\geq 3}$-imprecise}}\\ % & budget per file\\
      \hline
      1      & 2            & 2           & 71 \\         %  & 18  \\
      2      & 6            & 6           & 110\\         % & 22  \\
      4      & 3            & 5           & 158\\         % & 24, 89  \\
      8      & 10           & 10          & 193\\         % & 33  \\
      16     & 9            & 10          & 165\\         % & 23, 11  \\
      32     & \maxhl{11} & \maxhl{22} & \maxhl{201}\\% & 28, 40
      \hline
  \end{tabular}
}
\caption{Effect of batch-check size on the number of detected issues.}
\label{tab:BatchSize}
\end{table}

\textbf{RQ5: What is the effect of batch checks?}
To evaluate the influence of the batch-check size on the effectiveness
of our approach, we ran \tool on the same seed programs that were
selected for the evaluation of RQ4. We used the Uniform-Random search
strategy, and otherwise, the same configuration of our tool as in the
experiment of RQ4.

\tabref{tab:BatchSize} shows the effect of the batch-check size on the
number of detected issues.
Overall, larger batch-check sizes are more effective in detecting
soundness and precision issues. During our experiments, we also found
that larger batch-check sizes typically help in detecting the same
issues faster (that is, with a smaller initial budget) in comparison
to smaller sizes.

%% ----------------------------------------------------------------
\subsection{Threats to Validity}
\label{subsec:threats}
%% ----------------------------------------------------------------

We have identified the following threats to the validity of our
experiments.

\textbf{Selection of seed programs.}
Our experimental results may not generalize to other seed
programs~\cite{SiegmundSiegmund2015}. However, we evaluated our
technique by selecting seed programs from most categories of a
well-established repository of verification tasks~\cite{SV-COMP} and
by running the program analyzers on tens of thousands of program
variants. We, therefore, believe that our benchmark selection
significantly helps mitigate this threat and aids generalizability of
our results.

\textbf{Selection of program analyzers.}
For our experiments, we used the program analyzers described in
\secref{subsec:Analyzers}. Our findings depend on bugs, unsoundness,
and imprecision in these analyzers and, thus, may not
generalize. However, our selection includes a wide range of
program-analysis techniques, like model checking and abstract
interpretation. Moreover, all of these techniques are implemented in
mature tools.

\textbf{Type of checked properties.}
Our results may also not generalize to other types of checks, for
example, for points-to properties. Our implementation targets
numerical safety properties since they are found in almost every
program and can, therefore, be checked by most analyzers.
Independently, our approach and implementation are configurable and
may be extended to also synthesize other types of properties, for
instance, by checking if two pointers are aliases.

\textbf{Randomness in check synthesis.}
Another potential threat has to do with the internal
validity~\cite{SiegmundSiegmund2015} of our experiments, which refers
to whether any systematic errors are introduced in the experimental
setup. A typical threat to the internal validity of experiments with
randomized techniques is the selection of seeds. Recall that our
check-synthesis component selects candidate expressions and constants
in a randomized way. To ensure deterministic results and to avoid
favoring certain program analyzers over others, \tool uses the same,
pre-defined random seeds for all analyzer configurations.

\textbf{Survey of developers.}
A potential threat to the validity of our results is that the survey
questions were not understandable or presented in a clear way. To
alleviate this concern, we piloted the survey and fine-tuned the
questions based on the feedback we received.
Moreover, the survey responses might not be representative of other
professional C developers. However, we screened the candidates and
excluded any survey participants who did not seem experienced enough.

%% ----------------------------------------------------------------
\section{Related Work}
\label{sec:RelatedWork}
%% ----------------------------------------------------------------

In the literature, there are several techniques for evaluating
different qualities of program analyzers. Especially to ensure
soundness of an analyzer, existing work has explored a wide spectrum
of techniques requiring varying degrees of human effort, for instance,
manual proofs (e.g., \cite{MidtgaardAdams2012}), interactive and
automatic proofs (e.g., \cite{BessonCornilleau2013,BlazyLaporte2012}),
testing (e.g., \cite{MidtgaardMoeller2017,BugariuWuestholz2018}), and
``smoke checking''~\cite{BarnettChang2005}. There also exist
evaluations of the efficiency~\cite{SridharanFink2009} and
precision~\cite{LiangTripp2010} of various analyses.

Our approach is the first to differentially test real-world program
analyzers with the goal of detecting soundness and precision issues in
arbitrary code. Specifically, we identify such issues by comparing the
results of several analyzers, instead of relying on fixed test
oracles.

\textbf{Testing analyzers with randomly generated programs.}
Running a program analyzer on randomly generated input programs has
proved effective in revealing crashes~\cite{CuoqMonate2012}. However,
it is very challenging to randomly generate programs from scratch such
that they reveal soundness and precision issues in mature
analyzers. Instead, our approach takes as input existing, complex
programs as seeds and uses them to generate seed variants by
synthesizing checks for numerical properties.

\textbf{Testing symbolic execution engines.}
Kapus and Cadar use random program generation in combination with
differential testing to find bugs in symbolic execution
engines~\cite{KapusCadar2017}, by for instance comparing crashes,
output differences, and code coverage. Unlike our approach, this work
specifically targets symbolic execution engines and compares the tested
engines on randomly generated programs.

\textbf{Testing abstract interpreters.}
A common technique for revealing soundness issues in analyzers that
infer invariants (e.g., abstract
interpreters~\cite{CousotCousot1976,CousotCousot1977,CousotCousot1979})
is to turn inferred invariants into explicit assertions and then check if the assertions
are violated in concrete executions of the
program~\cite{CuoqMonate2012,WuHu2013,AndreasenMoeller2017}. Concrete
executions (e.g., from existing test suites) are also helpful in
identifying certain precision issues by observing the effect of
intersecting the inferred invariants with concrete runtime values on
the number of generated warnings. In contrast, our technique works for
any type of safety checker. In addition, if any of the
tested analyzers perform an under-approximation (e.g., a bounded model
checker or a dynamic symbolic execution engine), our technique
essentially compares the results of the other analyzers against a test
suite that is automatically generated on the fly.

A usual source of soundness and precision issues in abstract
interpreters is bugs in the implementation of the underlying abstract
domains and their operations (e.g., intersection and union of abstract
states). Existing
techniques~\cite{MidtgaardMoeller2017,MadsenLhotak2018,BugariuWuestholz2018}
for detecting such issues use well-known mathematical properties of
domains as test oracles. In contrast, our approach can not
only detect issues in domain implementations, but also in
abstract transformers, which model program statements such as
arithmetic operations or method calls.

\textbf{Evaluating unsoundness in static analyzers.}
Unsoundness is ubiquitous in static
analyzers~\cite{LivshitsSridharan2015}, typically to intentionally
favor other important qualities, such as precision or efficiency.
A recent technique systematically documents and evaluates the sources
of intentional unsoundness in a widely used, commercial static
analyzer~\cite{ChristakisMueller2015-Clousot}. The experimental
evaluation of this work sheds light on how often the unsoundness of
the analyzer causes it to miss bugs. In comparison, our approach
treats any tested program analyzer as a black box, and it is also able
to detect unintentional soundness and precision issues, caused by bugs
in the implementation of the analyzers.

Even more recently, an empirical study evaluated how many known bugs
are missed by three industrial-strength static bug
detectors~\cite{HabibPradel2018}. An important difference with our
approach is that the checked properties in this study did not
necessarily lie within the capabilities of the analyzers. In contrast,
we synthesize numerical properties, which should be handled by all
analyzers we tested. Moreover, our approach automatically synthesizes
potentially erroneous programs and uses differential testing to
identify both soundness and precision issues.

\textbf{Formally verifying program analyzers.}
To avoid any soundness issues in the first place, interactive theorem
provers are often used to verify the soundness of the \emph{design} of
a program analyzer. For instance, this is a common approach for type
systems~\cite{Dubois2000,ShaoSaha2002}. However, such proofs cannot
generally guarantee the absence of soundness issues in the actual
\emph{implementation} of the analyzer. To address this problem, the
Verasco~\cite{JourdanLaporte2015} project generates the executable
code of several abstract domains directly from their Coq
formalizations. Even if this approach were more practical, it would
still not easily detect precision issues in an analyzer.

\textbf{Testing compilers.}
Compilers typically apply different lightweight program analyses
(e.g., constant propagation) to produce more efficient code. Existing
work~\cite{YangChen2011,LeAfshari2014,LeSun2015,LidburyLascu2015,SunLe2016}
has proposed several techniques for detecting bugs in compilers, and
indirectly, in their analyses. These techniques often use metamorphic
testing to derive test oracles~\cite{BarrHarman2015}, for instance, by
comparing the output of two compiled programs where one is a slight,
semantics-preserving modification of the other. In contrast, our
approach compares several analyzers at once and uses their results to
detect soundness and precision issues. In addition, our check
synthesis instruments the seed program with assertions that may alter
its semantics, for example, by introducing failing executions.

%% ----------------------------------------------------------------
\section{Conclusion}
\label{sec:Conclusion}
%% ----------------------------------------------------------------

We have proposed a novel and automated technique for
differentially testing the soundness and precision of program
analyzers. We used it to test six mature, state-of-the-art analyzers
on tens of thousands of programs. Our technique found soundness and
precision issues in four of these analyzers.

In future work, we plan to explore how to synthesize checks for
different types of properties (for instance,
hyperproperties~\cite{ClarksonSchneider2008} like information flow,
and liveness properties like termination). We also plan to apply our
technique on a larger scale by using safety-critical programs, such as
flight controllers, as seed programs with the goal of generating new,
challenging verification benchmarks.

\section*{Acknowledgments}
We are very grateful to the program-analysis designers that we contacted, Matthias Dangl,
Arie Gurfinkel, Jorge Navas, and Zvonimir Rakamaric, for their prompt replies and fixes as
well as for their insightful explanations.

\newpage

%% Bibliography
\bibliographystyle{IEEEtran} \bibliography{tandem}

\end{document}